\documentclass[a4paper,fleqn,usenatbib]{mnras}

\usepackage[T1]{fontenc}
\usepackage{ae,aecompl}

\usepackage{graphicx}	
\usepackage{amsmath}	
\usepackage{amssymb}	

\title[Angular Momentum of Gas Rich Dwarf Galaxies]{Angular Momentum Content in Gas Rich Dwarf Galaxies}

\author[A. Chowdhury and J.N Chengalur]{
Aditya Chowdhury\thanks{E-mail: chowdhury@ncra.tifr.res.in}
and Jayaram N. Chengalur\thanks{E-mail: chengalur@ncra.tifr.res.in}
\\
National Centre for Radio Astrophysics, TIFR, Pune, India
}

\date{Accepted XXX. Received YYY; in original form ZZZ}

\pubyear{2015}

\begin{document}
\label{firstpage}
\pagerange{\pageref{firstpage}--\pageref{lastpage}}
\maketitle

\begin{abstract}

  We derive the specific baryonic angular momentum of five gas rich dwarf galaxies from HI kinematics complemented by stellar mass profiles. Since the gas mass of these galaxies is much larger than the stellar mass, the angular momentum can be determined with relatively little uncertainty arising from the uncertainties in the stellar mass to light ratio. We compare the relation between the specific baryonic angular momentum (j) and the total baryonic mass (M) for these galaxies with that found for spiral galaxies. Our combined sample explores the j-M plane over 3 orders of magnitude in baryon mass.  We find that our sample dwarf have significantly higher specific angular momentum than expected from the relation found for spiral galaxies. The probability that these gas rich dwarf galaxies follow the same relation as spirals is found to be $<10^{-6}$. This implies a difference in the evolution of angular momentum in these galaxies compared to larger ones. We suggest that this difference could arise due to one or more of the following : a lower baryon fraction in dwarf galaxies, particularly that arising from preferential outflows low angular momentum gas as found in high resolution simulations that include baryonic feedback;  "cold mode" anisotropic accretion from cosmic filaments. Our work reinforces the importance of the j-M plane in understanding the evolution of galaxies.

\end{abstract}




\section{Introduction}
$\\$
The mass and angular momentum of galaxies depend on their evolutionary history. Angular momentum is thought to have originated from tidal torquing in the primordial universe when proto-galaxies were collapsing \citep{ref:Peebles}. In hierarchical galaxy formation models, mergers play an important role. The mass and angular momentum evolution also depends on the merger history ('major' vs 'minor' mergers; 'wet' vs 'dry' mergers, 'cold' accretion etc). Finally, angular momentum is also correlated with the morphology. Systems with high angular momentum have thin disks while those with low angular momentum have bulges.

Since the angular momentum (J) depends on mass (M), it is useful to talk of the angular momentum per unit mass or the specific angular momentum defined as $j=J/M$. The dependence of this quantity (j) on the total mass of the galaxy was first studied observationally by \citet{ref:fall83} where he examined the location of spirals and ellipticals on the j-M plane. For both types, j was found to have a power law dependence on M : $j=q M^{\alpha}$ with the power law index $\alpha\approx2/3$. However, the scale factor q was found to be significantly lower in ellipticals indicating severe loss of angular momentum as they evolved. This work was further expanded by \citet{ref:rom_fall2012} and \citet{ref:rom_fall2013} in which a larger sample was looked at and it further reinforced the dependence of the mass-angular momentum relation on the Hubble type of the galaxy. They concluded that for a fixed mass, the Hubble type is a function of the specific angular momentum. Both of these studies derived the angular momentum from velocity width measurements and assumed a specific form for the rotation curve of the galaxies. 

\citet{ref:Obr2014} explored the j-M relation using kinematics of 16 spiral galaxies derived from the THINGS survey \citep{ref:things}. Instead of assuming a specific velocity curve, these authors performed an integration of $dJ$ over the galactic disk to find the total angular momentum J. They also quantified the morphological dependence of the j-M relation via bulge fraction ($\beta$) and found a tight correlation between the specific baryon angular momentum ($j_b$), the mass ($M_b$) and the bulge fraction ($\beta$). 

The j-M correlation can be qualitatively understood in CDM models where the angular momentum originates from tidal torquing in the early universe and is conserved to different extents in different evolution scenarios \citep{ref:rom_fall2012}. This argument is now supported by  results from N-body simulations that incorporate baryon feedback \citep{ref:bar_sim1,ref:bar_sim2}.  Without feedback, the angular momentum content produced in hydrodynamic N-body cosmological simulations is easily lost resulting in production of more bulge galaxies than observed \citep{ref:ang_mom_crisis1,ref:ang_mom_crisis2}. High resolution simulations indicate that feedback leads to outflows that preferentially remove low angular momentum material from the disk \citep{ref:sim_feedback,ref:sim_dwarf_zoom}. The efficiency of such processes naturally depends on the extent of dominance of feedback in the galaxy \citep{ref:sim_illustris}.

The angular momentum content of dwarfs have been looked at by \citet{ref:Bosch_2001} in connection to the distribution of specific angular momentum in such galaxies and how they compare to predictions from N-body hydrodyamic simulations. The angular momentum content of dwarfs with respect to their location on the j-M plane has remain unexplored. They play an important role in the hierarchical formation scenario where such dwarf galaxies merge to form larger galaxies. Their angular momentum content will provide insights into how it is retained in such an evolutionary scenario and the role of supernovae feedback in doing so. 

In this paper we look at the angular momentum content of a subclass of such dwarf galaxies which are very gas rich. The measurement of mass and angular momentum of these galaxies would be negligibly affected by uncertainties in the stellar mass-to-light ratio. We analyse the region in the j-M-$\beta$ space which they occupy and how it compares to the correlation obtained by \citet{ref:Obr2014}. In section \ref{sec:sample} we describe the galaxies analysed. Section \ref{sec:data_analysis} elaborates how the data were analysed to compute the total baryonc mass and the specific baryon angular momentum. In section \ref{sec:location} we discuss the angular momentum content of these galaxies with respect to the j-M-$\beta$ correlation. We interpret these results in \ref{sec:interpretation} and discuss why these dwarfs occupy a different region in the j-M plane. Finally we conclude in section \ref{sec:conclusion} and summarize the results obtained.

\section{Sample}
\label{sec:sample}
Our sample consists of all the known gas rich dwarf galaxies in the local universe with HI to stellar mass ratio greater than $~8$ for which high quality interferometric data are available. These galaxies, along with the telescopes used for the HI observations, are listed in Table~\ref{tab:galaxy_list}. The baryonic masses of these galaxies (along with other derived parameters) are listed in Table~\ref{tab:ang_mom}. As can be seen, all these galaxies have baryonic mass below $10^9 M_\odot$ and enables us to look at previously unexplored regions on the j-M plane.
\begin{table}
	\centering
	\label{tab:galaxy_list}
	\begin{tabular}{llcc} 
		\hline
		Galaxy & Telescope & Distance (Mpc) & $M_{gas}/M_{*}$\\
		\hline
		DDO 154 & VLA &4.04 $\pm$ 0.03 & 51\\
		DDO 133 & VLA &5.11 $\pm$ 0.09 &23\\
		NGC 3741 & WSRT &3.24 $\pm$ 0.05 &41\\
		UGCA 292 & GMRT &3.77 $\pm$ 0.06 &8 \\
		ANDROMEDA IV & GMRT &7.18 $\pm$ 0.14 &14\\
		\hline
	\end{tabular}
	\caption{List of galaxies that were analysed and their gas to stellar mass fraction. The masses used to derive the fraction are from this paper. The distances measurements, using fits to the tip of red giant branch (TRGB), are from Cosmicflows-2 \citep{ref:cosmicflow}, except DDO 154 which was derived in \citet{ref:dist_ddo154}. }
\end{table}
\section{Data Analysis}
\label{sec:data_analysis}
The interferometric data was flagged and calibrated in the Common Astronomy Software Applications (CASA, \citet{ref:casa}) package using standard procedure. Subsequent continuum subtraction and imaging were done using the same package. The integrated HI intensity maps (moment0) along with the velocity field (moment 1) for the  galaxies are shown in Appendix \ref{appn:summamry} . 

We used the Fully Automated TiRiFiC (FAT) pipeline developed by \citet{ref:fat} to fit these data cubes to tilted ring models. The pipeline uses TiRiFiC \citep{ref:tirific} to do a three dimensional fit to the data cube and is different from a traditional two dimensional tilted ring fit to the velocity field. The program provides the best fit HI surface brightness distribution as well as the rotation curve for each galaxy. These derived profiles are given in Appendix  \ref{appn:summamry}.

 Estimating the error on the best fit parameters from titled ring fits is non trivial. Traditionally either the velocity residual along a ring or the asymmetry between the approaching and receding side has been used a measure of uncertainty. In this paper we take the velocity residual along each ring as a pseudo-1sigma error on the rotational velocity. The residual velocity along a ring was calculated using the following relation:
 \begin{equation*}
 \sigma_v(r)=\frac{\sqrt{\sum_{r}^{r+\Delta r}{(v_{model}-v_{data})^2}}}{\sin{i(r)}}
  \end{equation*}
 where $\Delta r$ is the width of the ring at radius r; i(r) is the inclination of the same; $v_{model}$ and $v_{data}$ are calculated from the mom 1 map of the model and the data respectively. The uncertainty thus defined is a measure of the modelling residual along each ring. We similarly define psuedo-1sigma uncertainties for surface brightness (SBR) at each radii:
  \begin{equation*}  
  \sigma_{SBR}(r)=\cos{i(r)}\sqrt{\sum_{r}^{r+\Delta r}{(SBR_{model}-SBR_{data})^2}} 
  \end{equation*}
 where $SBR_{model}$ and $SBR_{data}$ are calculated from the mom 0 map of the model and the data respectively. Along with the residual error we also include a 10$\%$ flux calibration error in the surface brightness profile.  We note that while the errors thus defined are not \emph{formal errors}, they include errors from other parameters such inclination, position angle and central velocity and thus will tend to overestimate the physical errors. 
\subsection{Computation of Angular Momentum and Mass}
\label{ssec:ang_comp}
The computation of the total baryon angular momentum and total mass requires (i) the surface density profile of gas (ii) the surface density profile of stellar mass (iii) the rotation curve. The molecular gas content in dwarf galaxies have been observed to be small \citep{ref:mol_gas2,ref:mol_gas}. We thus neglect $H_2$ masses and obtain the distribution of gas from the HI surface brightness profile using a constant HI to Helium conversion ratio. From the measurements of Helium abundance in the local universe by \citet{ref:he_abundance} we take

\begin{equation*}
\frac{M_{HI}+M_{He}}{M_{HI}}=1.342\pm0.004
\end{equation*}

 Given the high gas to stellar mass ratio in our sample the steller mass distribution does not play a critical role in determining the total baryon angular momentum of the galaxy. In this work we use exponential fit parameters for the stellar mass distribution derived in the published literature to calculate the angular momentum and mass contained in stars.  For DDO 154, DDO 133 and UGCA 292 we use the exponential fits derived from IRAC 3.6$\mu$m images in \citet{ref:LittleThingsMass}. Similar I-band derived parameters for Andromeda IV were obtained from \citet{ref:mass_andiv}. And the fit parameters for NGC 3741 was taken from \citet{ref:n3741_stellar_scale,ref:n3741_stellar_mass}. All measurements of stellar scale length and mass from literature were corrected for difference in adopted distance. We note that the heterogeneity of mass to light ratio will have minimal effect given that the stellar mass makes up for less than $10\%$ of the total baryon mass in our samples. 

The total baryon mass ($M_b$) and the total angular momentum perpendicular to the disk ($J_b$) were computed by numerically evaluating the following integrals:

\begin{equation}
\label{eqn:tot_ang}
J(R)=\int_{0}^{R}{dr \ 2\pi r \  \Sigma_b(r) \ v(r)cos\left\lbrace\delta i(r)\right\rbrace \ r }
\end{equation}

\begin{equation}
\label{eqn:tot_mass}
M(R)=\int_{0}^{R}{dr \ 2\pi r \  \Sigma_b(r)}
\end{equation}

In equation (\ref{eqn:tot_ang}), $\delta i(r)$ is the difference in inclination with respect to the central disk inclination. The total baryon mass and angular momentum within radius R as a function of R can be found in Appendix \ref{appn:summamry}.   Table \ref{tab:ang_mom} lists the total mass and the angular momentum for stars and gas along with the error on these quantities. The same table also lists the baryonic specific angular momentum defined as  $j_b=J_b/M_b$.
\begin{table*}
	\centering
	\label{tab:ang_mom}
	\begin{tabular}{lccccccc} 
		\hline
		Galaxy &  $M_{*}$ & $M_{gas}$  & $M_{b}$ & $J_{*}$ &$J_{gas}$  & $J_{b}$ & $j_{b}$\\
	  	\ &  \multicolumn{3}{c}{ $\lg{M_\odot}$}  & \multicolumn{3}{c}{$\lg{M_\odot \ km/s \ pc}$}  &  $\lg{km/s \ pc} $ \\
		\hline
	DDO 154 & 7.16 & $8.64^{+0.07}_{-0.08}$ & $8.65^{+0.06}_{-0.08}$ & 11.90 & $13.91^{+0.09}_{-0.11}$ & $13.92^{+0.09}_{-0.11}$ & $2.27^{+0.08}_{-0.10}$\\ 
	DDO 133 & 7.63 & $8.48^{+0.09}_{-0.12}$ & $8.54^{+0.08}_{-0.10}$ & 12.70 & $13.54^{+0.11}_{-0.15}$ & $13.60^{+0.10}_{-0.13}$ & $2.06^{+0.09}_{-0.11}$\\ 
	NGC 3741 & 7.17 & $8.35^{+0.08}_{-0.10}$ & $8.37^{+0.08}_{-0.10}$ & 10.75 & $13.49^{+0.11}_{-0.15}$ & $13.49^{+0.11}_{-0.15}$ & $2.11^{+0.10}_{-0.13}$\\ 
	UGCA 292 & 6.32 & $7.76^{+0.11}_{-0.14}$ & $7.78^{+0.10}_{-0.13}$ & 10.87 & $12.08^{+0.15}_{-0.23}$ & $12.10^{+0.14}_{-0.21}$ & $1.33^{+0.13}_{-0.19}$\\ 
	AND IV & 7.39 & $8.53^{+0.07}_{-0.09}$ & $8.56^{+0.07}_{-0.08}$ & 11.87 & $13.72^{+0.09}_{-0.11}$ & $13.73^{+0.09}_{-0.11}$ & $2.17^{+0.07}_{-0.09}$\\ 
		\hline
	\end{tabular}
	\caption{Summary of quantities derived in this paper. Listed above, for each galaxy, is the stellar mass ($M_*$), gas mass ($M_{gas}$), total baryon mass ($M_b$); stellar angular momentum ($J_*$), gas angular momentum ($J_{gas}$) and total baryon angular momentum ($J_b$); and the baryon specific angular momentum ($j_b$) along with errors on them as computed using techniques described in section \ref{ssec:ang_comp} }
\end{table*}
\section{Location of gas rich dwarfs in the $j-M-\beta$ relation}
\label{sec:location}
\begin{figure}
 \includegraphics[width=\columnwidth]{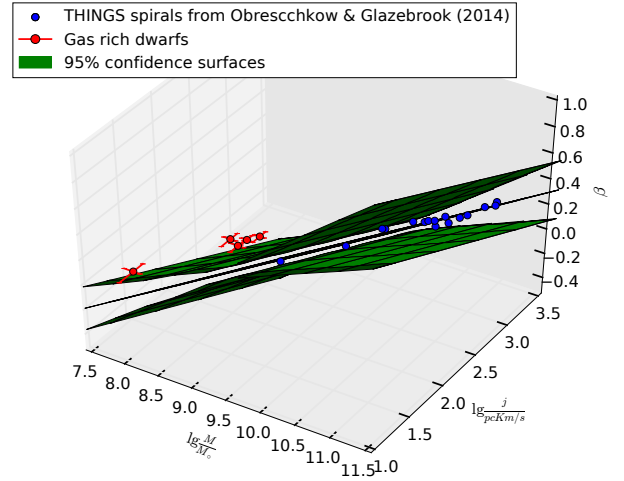}
 \caption{A view of the j-M-$\beta$ space. The figure shows the mean plane (given by equations \ref{eqn:obr_rel} $\&$ \ref{eqn:boot_best_fit}) surrounded by the $95.4\%$ surfaces as computed from the distribution of $(k_1,k_2,k_3)$ obtained in the bootstrap resampling procedure. The points derived in this paper for gas rich dwarfs are all outside the confidence space. Also shown are the points derived in \citet{ref:Obr2014}, all of which are well inside the $95.4\%$ space.} 
 \label{fig:3D_relation}
\end{figure}
\begin{figure} \includegraphics[width=\columnwidth]{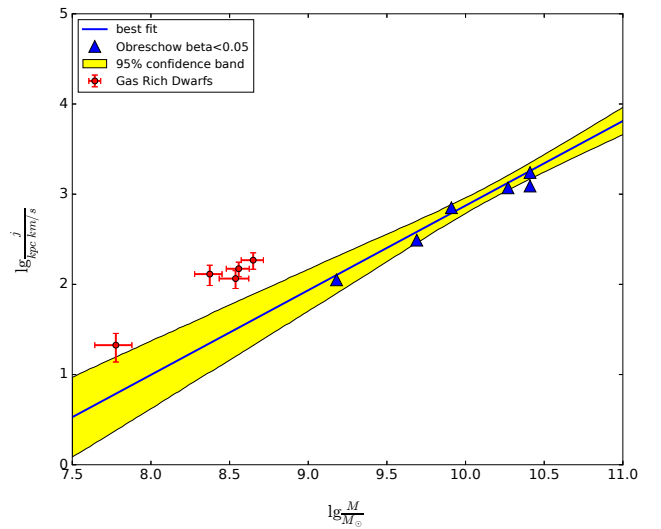}
 \caption{The $\beta=0$ plane of relation \ref{eqn:obr_rel_changed} along with the $95.4\%$ confidence interval inferred from the distribution of $(c_1,c_2,c_3)$. Note that the all the gas rich galaxies from this paper lie outside the confidence band. Also plotted for comparison are the low bulge fraction ($\beta<0.05$) samples from \citet{ref:Obr2014}.}
 \label{fig:2D_relation}
\end{figure}
\subsection{j-M-$\beta$ for Spiral Galaxies}
\citet{ref:Obr2014} analysed 16 spiral galaxies from the THINGS sample \citep{ref:things} and showed that the logarithm of mass ($M_b$), the logarithm of specific angular momentum ($j_b$) and the bulge fraction ($\beta$) are tightly correlated and lie in a plane in the j-M-$\beta$ space. They measure the mass and angular momentum from the full kinematics (as in this paper using eqn. (\ref{eqn:tot_ang}) and eqn. (\ref{eqn:tot_mass})). The details of fitting are slightly different - they used inclinations derived in \citet{ref:leroy} and assumed a flat disk while deriving the rotation curves from the velocity field whereas this paper fits a model to the three dimensional data cube. They calculated the bulge fraction by decomposing the stellar disk into an exponential disk and bulge described by the Sersic profile.

They define an empirical relation between $j_b$,$M_b$ and $\beta$ given by the equation below.  

\begin{equation}
\label{eqn:obr_rel}
\beta = k_1 \lg\left[\frac{M_b}{10^{10} M\circ}\right] + k_2 \lg\left[\frac{j_b}{10^3 \ kpc \ km \ s^{-1}}\right] + k_3
\end{equation}

The best fit values for the parameters in eqn. (\ref{eqn:obr_rel}) are given by \citet{ref:Obr2014} to be:
\begin{equation}
\label{eqn:obr_best_fit}
(k_1,k_2,k_3) = (0.34 \pm 0.03, -0.35 \pm 0.04, -0.04 \pm 0.02)
\end{equation}
\subsection{Confidence Bands in the Relation}

As mentioned above, we use as a reference, study of the 16 spiral galaxies by \citet{ref:Obr2014}, where a relation was derived between the the specific angular momentum, the mass and the bulge fraction. Our small gas rich galaxies do not have any stellar bulge. As such we would prefer to use the specific angular momentum $j_b$ rather than the bulge fraction $\beta$ as the independent variable in defining the relation between these three parameters. For this reason, and also in order to get an independent estimate of the errors in the underlying relation, we re-derive the errors in eqn. (\ref{eqn:obr_rel}) using the data from \citet{ref:Obr2014} and bootstrap resampling. 

Bootstrap re-sampling involves drawing samples of the dataset with replacement and then doing a chi-square minimization of the drawn sample to determine $(k_1,k_2,k_3)$. The above procedure was repeated $\approx 10^7$ times to obtain the distribution of each of the parameters. The most probable
parameters along with the marginalized $68\%$ errors are:

\begin{equation}
\label{eqn:boot_best_fit}
(k_1,k_2,k_3) = (0.33 \pm 0.04, -0.35 \pm 0.06, -0.03 \pm 0.02)
\end{equation}

The parameters that we get are broadly consistent with eqn. (\ref{eqn:obr_best_fit}) and hence does not affect conclusions drawn about the mean correlation. The bootstrap resampling procedure however gives slightly larger errors than that estimated by \citet{ref:Obr2014}. For consistency we derive the confidence band numerically using the complete distribution of $(k_1,k_2,k_3)$ from the bootstrap re-sampling procedure. The mean relation as defined by equation (\ref{eqn:obr_rel}) and (\ref{eqn:boot_best_fit}) along with 95.4$\%$ confidence bands is shown in Fig. \ref{fig:3D_relation}.  Also shown are the points derived in this paper from Table \ref{tab:ang_mom}.

As mentioned above,  all our gas rich dwarfs are bulge-less, i.e $\beta=0$, and it is hence more useful to change the dependent variable in equation (\ref{eqn:obr_rel}) to $\lg{j_b}$:
 \begin{equation}
 \label{eqn:obr_rel_changed}
\lg\left[\frac{j_b}{10^3 \ kpc \ km \ s^{-1}}\right]  = c_1 \lg\left[\frac{M_b}{10^{10} M\circ}\right] +  c_2 \beta + c_3
 \end{equation}
 
A similar bootstrap re-sampling with this as the fitting function gives the best fit parameters as:
 
 \begin{equation}
 \label{eqn:cboot_best_fit}
 (c_1,c_2,c_3) = (0.94 \pm 0.05, -2.56 \pm 0.24, -0.13 \pm 0.04)
 \end{equation}
  
 The $\beta=0$ plane, along with the 95.4$\%$ confidence band derived from the distribution of $(c_1,c_2,c_3)$ is shown in Figure \ref{fig:2D_relation}.
\subsection{Location of the dwarfs in the j-M plane}
\label{ssec:diff_pop}

From Fig. \ref{fig:2D_relation}, it is evident that the gas rich dwarfs \emph{all} lie above the mean j-M-$\beta$ relation. This is unlike the situation for very small bulge fraction  ($\beta < 0.05$) spirals from \citet{ref:Obr2014}, all of which do lie on the relation. Of even more significance is the fact that \emph{all} of the dwarfs lie outside the $95\%$ confidence band. This is clearly indicative of the fact that the angular momentum and mass of these galaxies do not follow the same relation as that of spirals. Our bootstrap resampling analysis also enables us to give an estimate of the probability ($P_{similar}$) that these dwarf galaxy data-points are derived from the relation followed by spirals. This is given by fraction of bootstrap instances when the best fit plane intercepts all five of the data-points within boxes defined by the errors in $j_b$ and $M_b$. The fraction is  $P_{similar}<10^{-6}$, indicating a negligible probablity that the dwarf galaxies follow the same trend as the spirals. In an recent independent analysis based on dwarfs from the LITTLE THINGS sample \citet{ref:butler16} come  to a similar conclusion.

\section{Interpretation}
\label{sec:interpretation}
The j-M relation obtained by \citet{ref:rom_fall2012} was explained in their work using basic CDM models along with tidal torquing in the early universe. Similar arguments were also used by \cite{ref:Obr2014} to explain their relation. We review and build upon these arguments and then interpret the result of the current work.
\subsection{Origin of the j-M scaling}

The asymmetric growth of perturbation in the primordial universe lead to tidal torques acting between proto-galaxies as they were collapsing to form early galaxies \citep{ref:Peebles}. This process is thought to continue till  turnaround. During this period angular momentum in proto-galaxies grows linearly with cosmic time and this leads to the origin of angular momentum in dark matter halos. In such a scenario, theory explicitly predicts the specific angular momentum of the dark matter halo ($j_H$) to be related to the mass of the halo ($M_H$) as \citep{ref:tidal_shaya}:
\begin{equation}
\label{eqn:ang_mass_scaling}
j_H = k {M_H}^{2/3}
\end{equation}
Where k is a constant that depends on the details of the tidal torquing process.

The energy of a virialized halo ($E_H$) at the current epoch is given by:
\begin{equation}
\label{eqn:energy_mass_scaling}
|E_H|= s \left(\frac{25 H_0^2 G^2}{8}\right)^{1/3} {M_H}^{5/3} 
\end{equation}
where $H_0$ is the present Hubble Constant and s depends on the kind of halo profile, for a singular isothermal halo it is unity and for a NFW profile it depends on the concentration parameter c.

The mass dependences in eqn. (\ref{eqn:ang_mass_scaling}) and  (\ref{eqn:energy_mass_scaling}) motivates a dimensionless spin parameter, first introduced by \citet{ref:Peebles}, that does not depend on the  final (post-torquing) halo mass ($M_H$) and angular momentum ($J_H$) :
\begin{equation}
\label{eqn:spin_param}
\lambda = \frac{J_{H} {|E_{H}|}^{1/2}}{G M_{H}^{5/2}} 
\end{equation}
Using equations \ref{eqn:ang_mass_scaling}-\ref{eqn:spin_param} we obtain :
\begin{equation}
\label{eqn:spin_param_const}
k(\lambda)  = \left(\frac{8}{25}\right)^{1/3}  \frac{s}{H_0^{2/3} G^{-1/3}} \  \lambda
\end{equation}
Thus given a cosmology, $\lambda$ essentially captures the details of the tidal torquing process and lets us evaluate $k$. And from eqn. (\ref{eqn:spin_param}) the spin parameter ($\lambda$) should be scale invariant and not depend on the mass of the galaxy. Indeed N-body simulations show that $\lambda$ of the dark matter halo follows a log-normal distribution which is independent of mass and environment \citep{ref:spin_param_bullock,ref:spin_param_mass_dep,ref:spin_param_mass_bett}. This implies that the initial angular momentum of the halo and its mass would follow eqn. (\ref{eqn:ang_mass_scaling}) with the scatter being dictated by the distribution of $\lambda$.

At early epochs we can expect the gas to be mixed with dark matter and thus obtain the same specific angular momentum. This enables us to write a similar relation between baryon angular momentum at that epoch ($j_b^0$) and the mass of the halo ($M_H$):
\begin{equation}
\label{eqn:b0_ang_mass_scaling}
j_b^0 = j_H = k(\lambda) {M_H}^{2/3}
\end{equation}

In the currently understood scenario, galaxies undergo non-linear processes such as mergers, outflows etc as they evolve. We note that eqn. (\ref{eqn:b0_ang_mass_scaling}) might not hold in cases where the galaxy has undergone complex mergers or outflows resulting in a large fraction of dark matter and  gas that makes up the galaxy at the present epoch to have been segregated in the protogalaxy stage \citep{ref:mass_segregation}.  In cases where the fraction of such segregation is small, the change in baryon mass and baryon angular momentum by these evolutionary processes can be parametrized by two dimensionless quantities, the angular momentum retention fraction : {$f_j=j_b/j_b^0$} and the baryon to dark matter fraction : $f_b=M_b/M_H$. These definitions allow equation (\ref{eqn:b0_ang_mass_scaling}) to be written as:
\begin{equation}
\label{eqn:b_ang_mass_scaling}
j_b = f_j \ f_b^{-2/3} \ k(\lambda) \ {M_b}^{2/3}
\end{equation}

 Note that both $f_j$ and $f_b$ may depend on the baryon mass of the galaxy and thus alter the scaling relation. Thus the observed difference between the scaling relations of different galaxy types will arise from the difference in the quantity $ f_j f_b^{-2/3}$ which in turn will depend on the complex evolution history of galaxies. This is discussed further by \citet{ref:rom_fall2012} while explaining the difference in spirals and ellipticals as seen on the j-M plane. 

\subsection{Explaining the Location of Dwarfs on the j-M Plane}

The significantly higher angular momentum content of gas rich dwarf galaxies implies a higher value of $<f_j f_b^{-2/3}>$ in eqn. (\ref{eqn:ang_mass_scaling}). Thus compared to spirals, our sample of galaxies have either a lower baryon fraction or a higher angular momentum retention fraction or a combination of both that leads to an increase in the average value of $f_j f_b^{-2/3}$. We discuss three possible mechanisms that can lead to the above.

\begin{itemize}
\item The baryon fraction $f_b$ of a galaxy as a function of its halo mass is expected to decrease with halo mass \citep{ref:baryon_mass_1,ref:baryon_mass_2}. This could be due to a variety of reasons including escape of baryons from the shallow dwarf potential wells during reheating and reionisation as well as escape via supernovae feedback as discussed further below. This implies a higher  $<f_j f_b^{-2/3}>$ for our sample of low baryon mass galaxies if we assume $f_j$ to be independent of halo mass.

\item  Further, N-body simulations that incorporate baryon feedback have shown that outflows driven by feedback are biased towards removing low angular momentum material leading to an enhancement of the specific angular momentum content of the galaxy \citep{ref:sim_feedback,ref:sim_feedback_2}. \citet{ref:sim_dwarf_zoom} did a high resolution, supernovae feedback incorporated, zoom simulation of a dwarf galaxy and showed that supernovae driven outflows effectively removes the low specific angular momentum gas in the inner part of the halo.  In dwarf galaxies which have shallow potential wells, supernovae can be very efficient in driving gas off the galactic disk. This explanation is further supported by the work of \cite{ref:Bosch_2001} where the authors looked at the specific angular momentum distribution in dwarfs and found a small fraction of baryons to occupy the lower specific angular momentum end of this distribution than predicted by hydrodynamic N-body simulations. This hints at mechanisms at work for dwarf galaxies which preferentially remove low specific angular momentum material from the galactic disk.

\item A key mechanism by which the angular momentum content of galaxies evolve is gas accretion. \citet{ref:gas_acc_keres} showed that two kinds of accretion operate in the galaxy formation epoch - (i) "Hot mode" where gas from filaments shock heat to the halo's virial temperature before collapsing to the galaxy. (ii) "Cold mode" where the gas falls into the galaxy before reaching the virial temperature. They find "Hot mode" accretion to be isotropic and "Cold mode" to be directed along filaments and hence anisotropic. In the same work, the latter is shown to dominate in galaxies with mass $M_b < 10^{10.3} M_\odot$. A large number of numerical studies have demonstrated the high  specific angular momentum content of "cold-mode" accreting gas (e.g \citet{ref:cold_mode_1,ref:cold_mode_2,ref:cold_mode_3,ref:cold_mode_4}). Thus it is possible that the sample of gas rich galaxies studied in this paper had experienced dominating "cold-mode" accretion in the early stages of their evolution leading to a higher specific angular momentum content. Also, much like cold mode accretion, wet mergers (particularly with a
satellite that was originally in a prograde orbit) could lead to a disk with higher specific angular momentum. One such interesting system has been recently found where a dwarf galaxy disk appears to be forming out of the merger of 3 smaller progenitors flowing along a filament \citep{ref:U3672_2015}.
\end{itemize}

While with the existing data we cannot come to a firm conclusion as to
which (if either!) of these two mechanisms is dominant, the fact that there
exist plausible mechanisms that would lead to a higher specific angular momentum in dwarf galaxies is interesting.

\section{Conclusions}
\label{sec:conclusion}

In this paper we present the measurements of the angular momentum content in dwarf galaxies using full HI kinematics. Our sample galaxies are gas dominated, which means that the uncertainties in the stellar mass to light ratio do not qualitatively affect our conclusions. The core finding of work is that we find,
to a very high level of significance that the gas rich dwarf galaxies are not drawn from the same $j-M$ correlation as spiral galaxies. For a given mass, the dwarfs have more baryon specific angular momentum than predicted from the existing correlation between mass and angular momentum in large galaxies.  The larger specific angular momentum contained in these galaxies can be due to (i) the lower baryon fraction in these galaxies, or perhaps more particularly the supernovae outflow driven removal of low angular momentum gas, (ii) "cold mode" accretion from cosmic filaments or some combination of these two. These findings, combined with existing observational studies \citep{ref:fall83,ref:rom_fall2012,ref:rom_fall2013,ref:Obr2014},  reinforces the view that the specific angular momentum  is a useful metric in understanding galaxy evolution.

\section*{Acknowledgements}
We would like to thank NCRA, NRAO and ASTRON for making public archival data taken with the GMRT, VLA and WSRT respectively. We are grateful to Peter Kamphuis for many discussions on fitting tilted ring models with his automated pipeline. He also provided useful inputs on how to extract errors on the velocity and SBR measurements. We also thank the anonymous referee for some helpful comments on the paper.




\bibliographystyle{mnras}
\bibliography{bibliography} 



\appendix

\section{A Look at Individual Galaxies}
\label{appn:summamry}
In this section we show for each galaxy in our sample (a) The mom 0 (total intensity) map, (b) The moment 1 (velocity field) map - The contours on this map are spaced 10km/s apart, (c) The surface density profile for both gas and stars, (d) The rotation curve (e) The total angular momentum within radius r as a function of r for both gas and stars (f) The mass enclosed within radius r as a function of r for both gas and stars . The description of how (a),(b),(c) $\&$ (d) were obtained is given in section \ref{sec:data_analysis} and the same for (e) $\&$ (f) in section \ref{ssec:ang_comp}.
\begin{figure*}
	\centering
 \includegraphics[width=2\columnwidth]{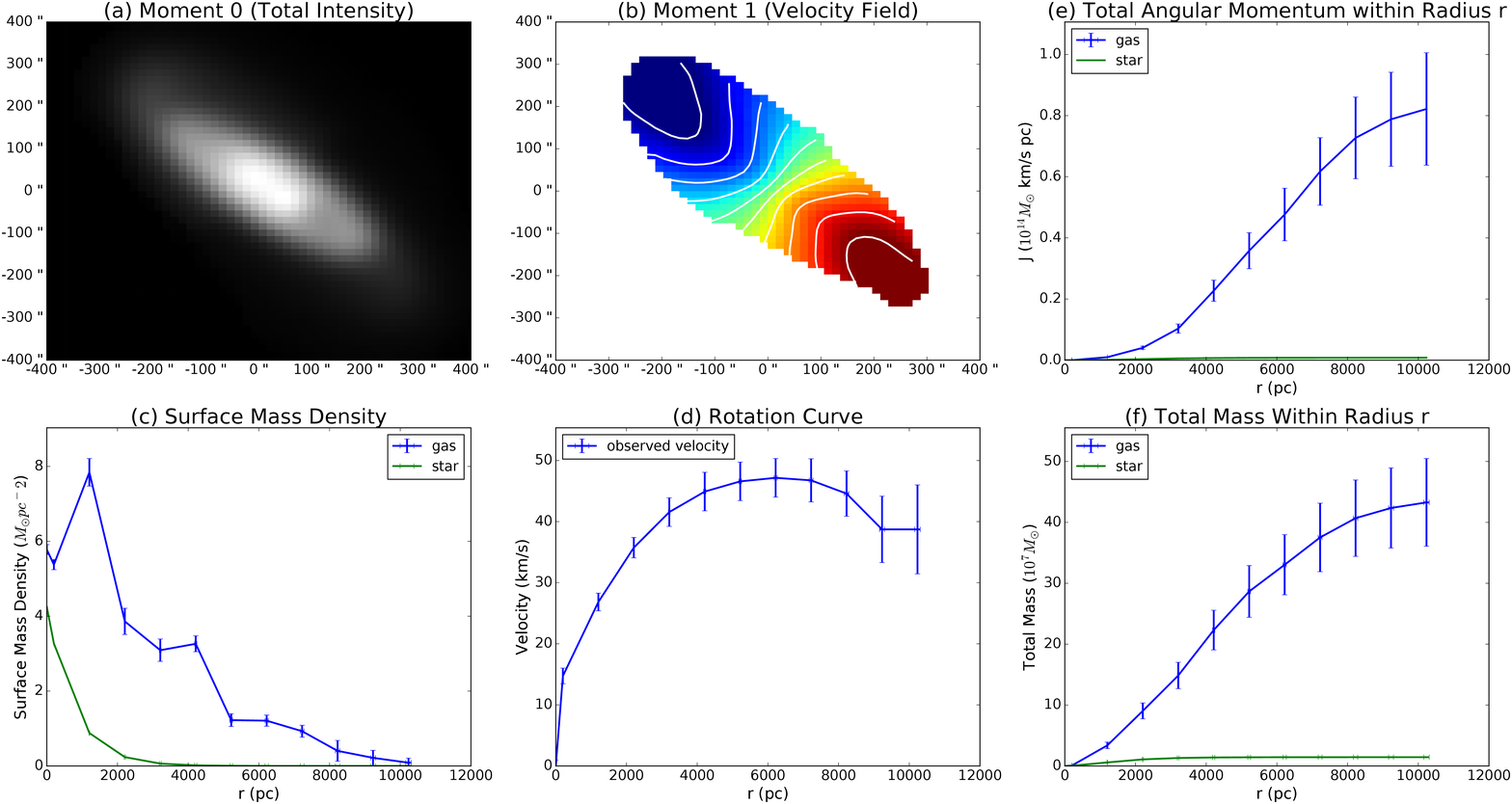}
 \caption{Overview of measurements for galaxy DDO 154. For description of the plots see introductory comments in Appendix \ref{appn:summamry}. The galaxy has a warp and its inclination  varies between 50$^\circ$ to 70$^\circ$. The systematic velocity of the galaxy centre is 352.0 km/s. Note that though there is an apparent drop in the rotation velocity at the outer edge, its consistent with a flat rotation curve within the errors. }
 \label{fig:overview_ddo_154}
\end{figure*}

\begin{figure*}
	\centering
 \includegraphics[width=2\columnwidth]{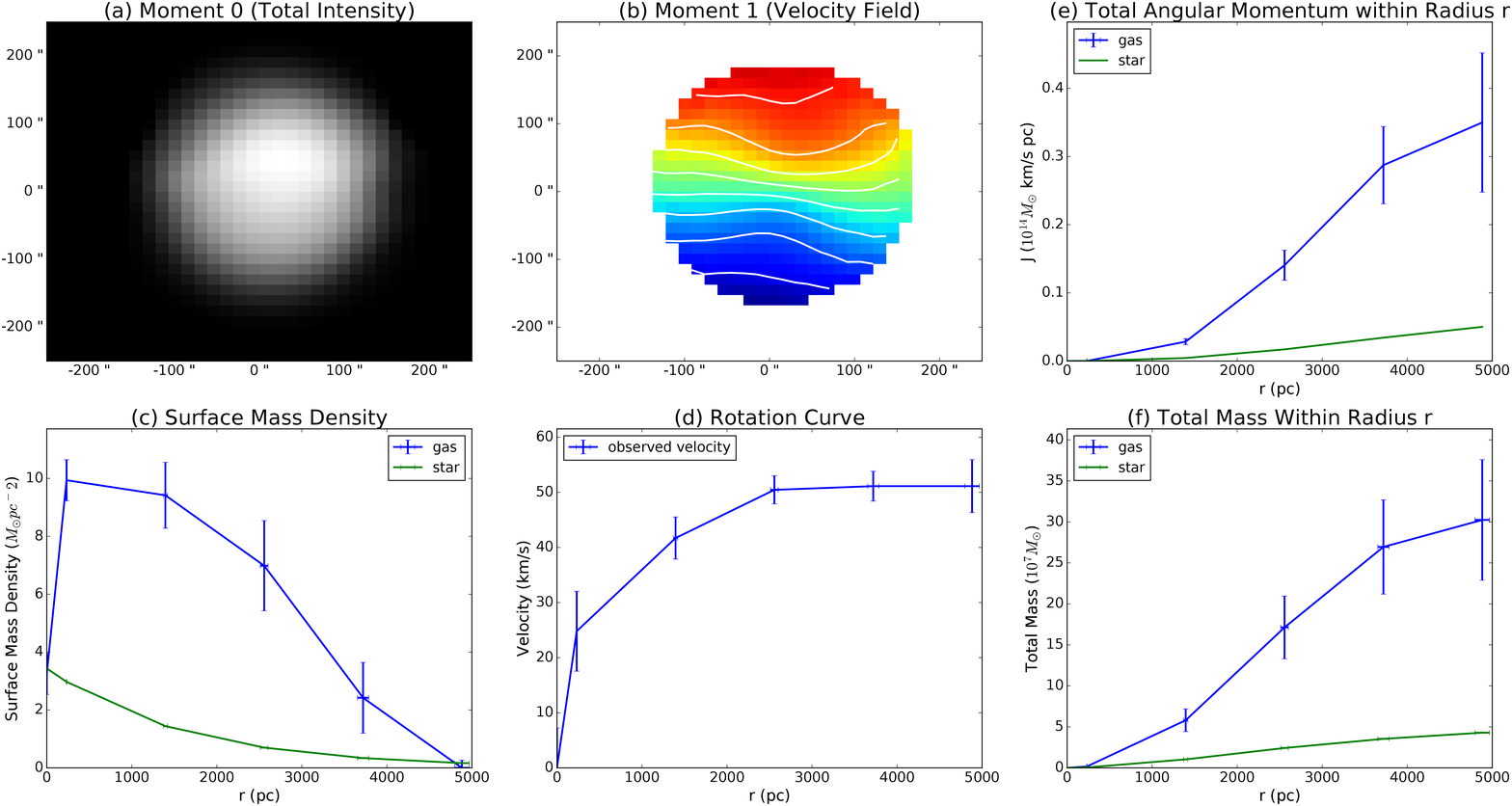}
 \caption{Overview of measurements for galaxy DDO 133. For description of the plots see introductory comments in Appendix \ref{appn:summamry}. The galaxy has a low inclination of ~29$^\circ$. The systematic velocity of the galaxy centre is 330.2 km/s.}
 \label{fig:overview_ddo_133}
\end{figure*}

\begin{figure*}
	\centering
 \includegraphics[width=2\columnwidth]{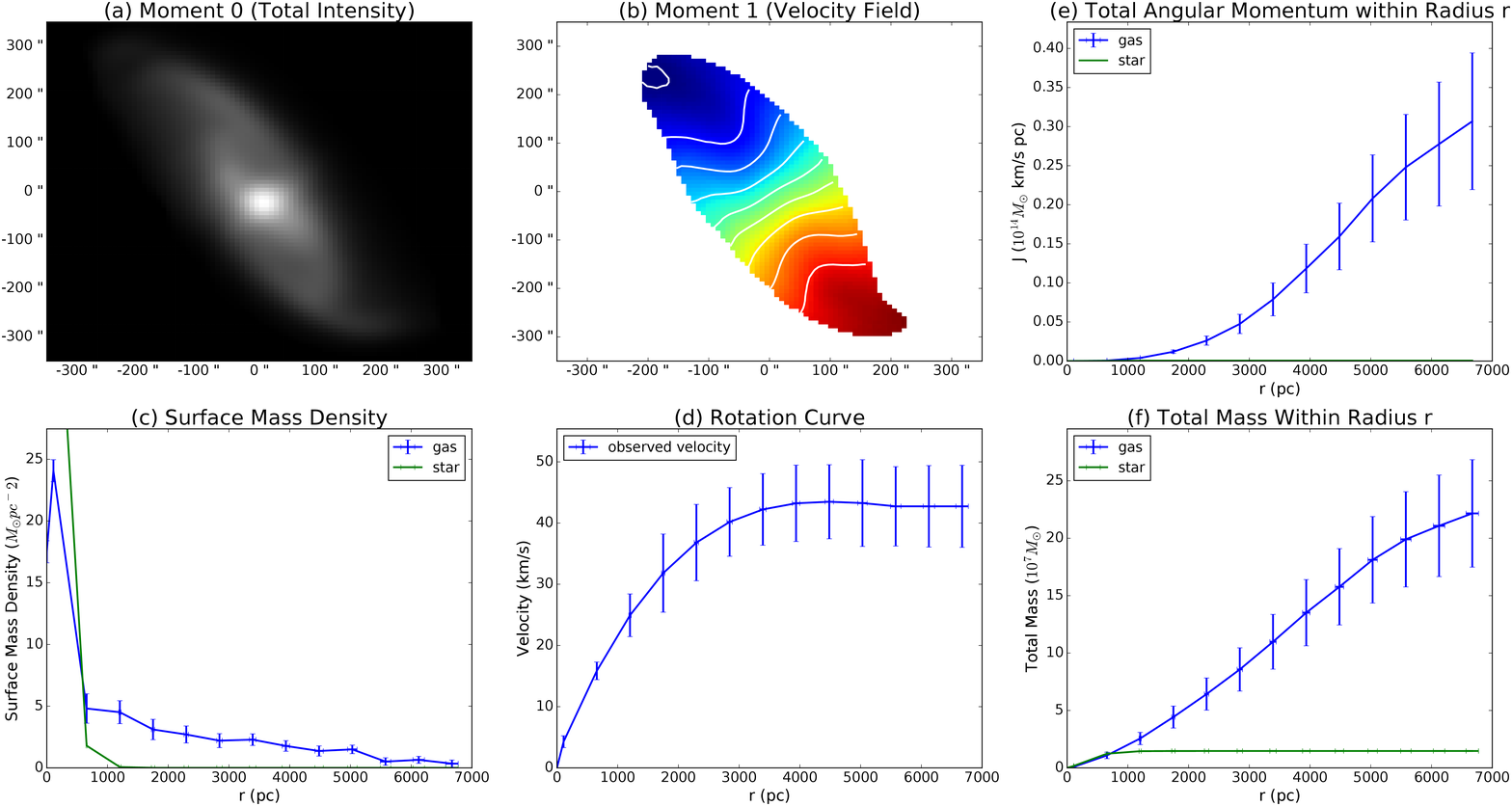}
 \caption{Overview of measurements for galaxy NGC 3741. For description of the plots see introductory comments in Appendix \ref{appn:summamry}. The galaxy has a warp and its inclination varies between 65$^\circ$ to 75$^\circ$. The systematic velocity of the galaxy centre is 223.7 km/s.}
\end{figure*}
\begin{figure*}
	\centering
 \includegraphics[width=2\columnwidth]{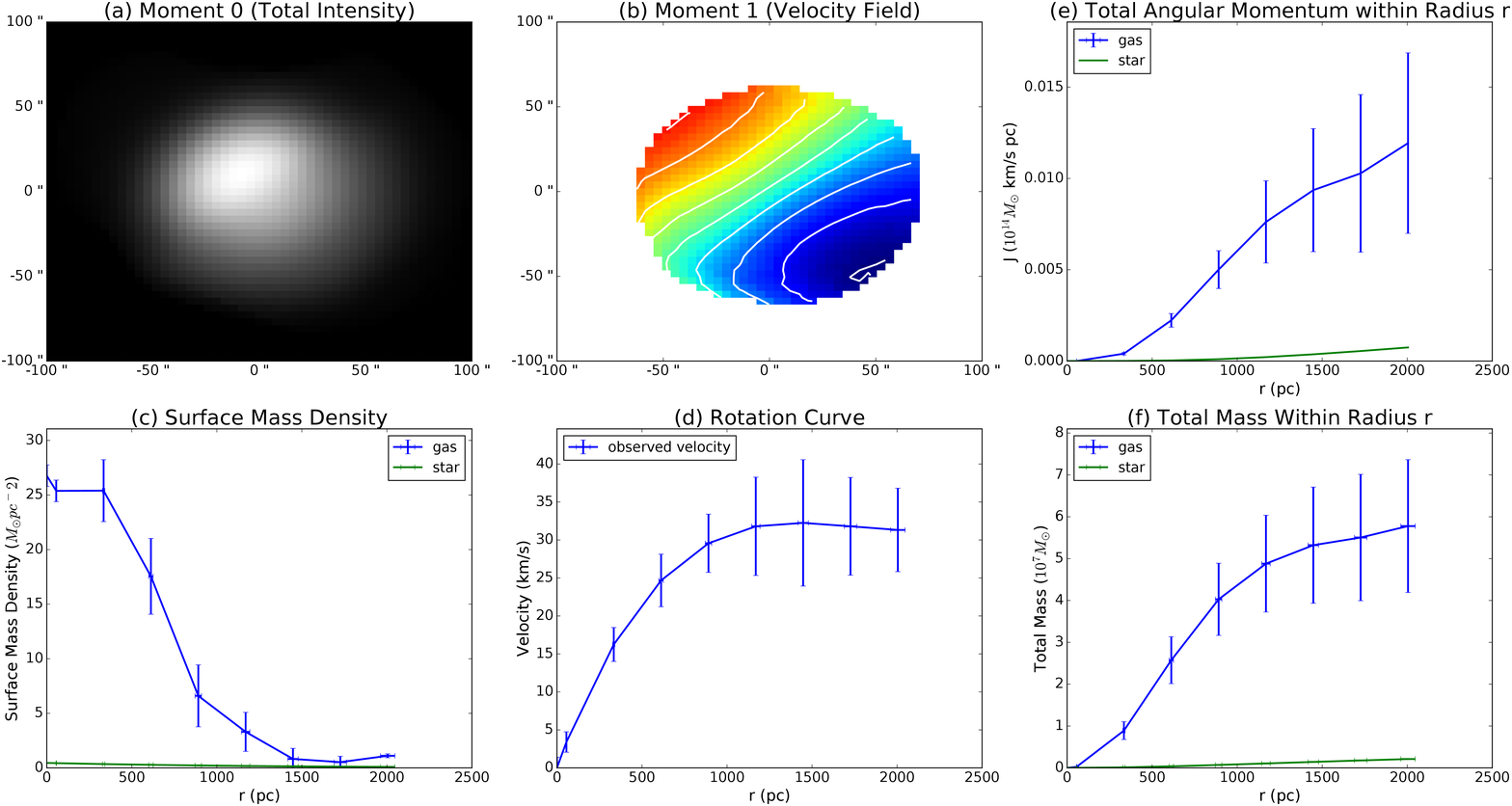}
\caption{Overview of measurements for galaxy UGCA 292. For description of the plots see introductory comments in Appendix \ref{appn:summamry}. The galaxy has a low inclination of ~19$^\circ$. The systematic velocity of the galaxy centre is 282.4 km/s.}
\end{figure*}
\begin{figure*}
	\centering
 \includegraphics[width=2\columnwidth]{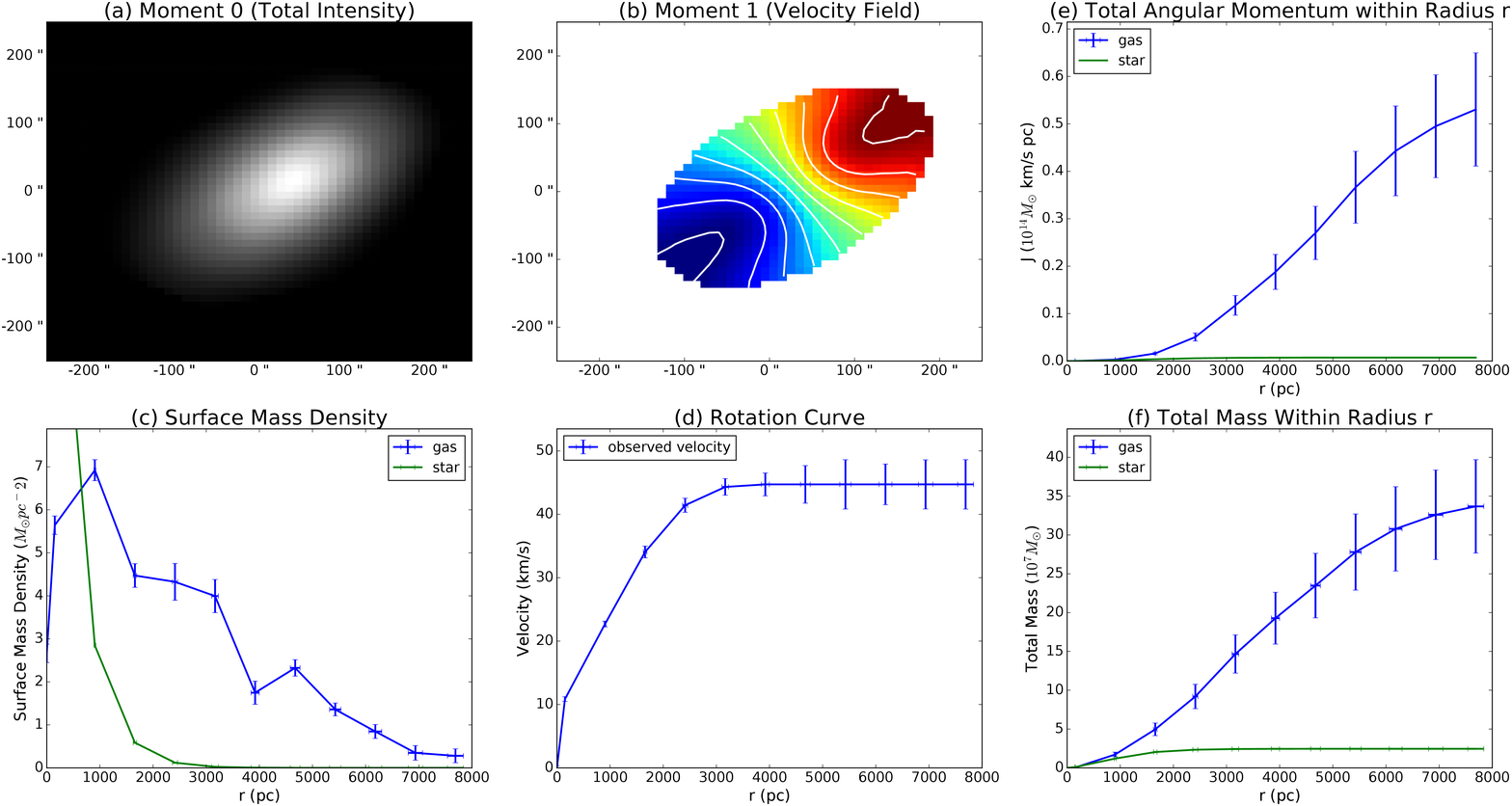}
 \caption{Overview of measurements for galaxy Andromeda IV. For description of the plots see introductory comments in Appendix \ref{appn:summamry}. The galaxy has an inclination of ~55$^\circ$ with no visible sign of warps. The systematic velocity of the galaxy centre is 256.8 km/s.}
\end{figure*}


\bsp	
\label{lastpage}
\end{document}